\documentclass{iopjournal}

\usepackage[english]{babel}
\usepackage{graphics}
\usepackage{epsf}
\usepackage{epsfig}
\usepackage{booktabs}
\usepackage{amsmath}
\usepackage{amsfonts}
\usepackage{amssymb}
\usepackage{graphicx}
\usepackage{color}
\usepackage{bm}
\usepackage[yyyymmdd]{datetime}
\usepackage{dashrule}
\usepackage{ulem}

\newcommand{\SIOx}{SrIr$_{1-x}$Sn$_{x}$O$_3$}
\newcommand{\SIO}{SrIrO$_3$}

\newcommand{\SIOe}{SrIr$_{0.9}$Sn$_{0.1}$O$_3$}
\newcommand{\SIOr}{SrIr$_{0.8}$Sn$_{0.2}$O$_3$}

\begin{document}

\articletype{Letter}

\title{Evolution of effective magnetic exchange interaction under spin dilution in \SIOx{}}

\author{Xiang Li$^1$\orcid{0009-0002-1677-6703}, Yifan Jiang$^1$\orcid{0000-0002-1477-4731}, Yuan Wan$^2$\orcid{0000-0001-8132-6691} and Xuerong Liu$^{1,3,*}$\orcid{0000-0001-7688-1156}}

\affil{$^1$School of Physical Science and Technology, ShanghaiTech University, Shanghai, China}

\affil{$^2$Institute of Physics, Chinese Academy of Sciences, Beijing, China}

\affil{$^3$Center for Transformative Science, ShanghaiTech University, Shanghai, China}

\affil{$^*$Author to whom any correspondence should be addressed.}

\email{liuxr@shanghaitech.edu.cn}


\begin{abstract}

Resonant inelastic X-ray scattering measurements reveal robust magnetic excitations in the perovskite iridates \SIOx{}. We analyzed the dispersions of the magnetic excitation with spin-dilution ratio $x$ = 0, 0.03, 0.06, 0.1, and 0.2, crossing from semi-metal to spin-diluted while antiferromagnetically ordered insulators. The extracted effective magnetic exchange interactions decrease continuously upon increasing spin dilution, and their evolution follows a simple spin-dilution scaling law. These results not only verify the strong electron-correlation nature of the metallic parent \SIO, but also reveal the entanglement of the charge and spin dynamics in this system. 

\end{abstract}

\section{Introduction}

Nonmagnetic substitution provides an effective route to tune the spin dynamics in materials and to drive emergent phenomena \cite{RevModPhys.81.45}, such as spin dilution induced suppression of superconductivity \cite{PhysRevLett.91.067002,PhysRevLett.105.037207} and the emergence of spin-glass behavior in frustrated magnetic systems \cite{PhysRevLett.64.2070,PhysRevB.89.241102}. In an antiferromagnetic (AFM) ordered system, a direct consequence of spin dilution is the reduction of magnetic ordering temperature, eventually leading to the breakdown of long-range AFM order after reaching the percolation threshold \cite{PhysRevB.65.104407,PhysRevB.66.024418}. This behavior has been widely observed in the Zn doped parent compounds of cuprate superconductors \cite{vajk2002quantum,PhysRevB.44.9739,PhysRevB.45.7430,PhysRevB.48.3485}, serving as a probe to explore the relation between the superconductivity and the spin dynamics. The magnetic moments are found to be preserved in the dilution process \cite{vajk2003neutron,Horsley_2022,JPSJ.91.014801}, while the magnetic ordering temperature is almost linearly decreasing under spin dilution at low dilution region \cite{PhysRevB.44.9739,PhysRevB.45.7430}. Thus the local exchange driven by strong electron correlation is still a proper physical picture.

Another route to suppress the AFM ordering is by introducing charge carriers, which strongly entangle the charge and spin dynamics and lead to competition between the local and itinerant magnetism \cite{RevModPhys.78.17,PhysRev.147.392}. On the other hand, if nonmagnetic impurities are further introduced, they could suppress the motion of charge carriers and instead help restore the long-range magnetic order \cite{RevModPhys.81.45}, as observed in Zn doped superconducting cuprates La$_{1-x}$[Sr/Ba]$_x$[Cu$_{1-y}$Zn$_y$]O$_4$ \cite{PhysRevLett.91.067002,PhysRevB.103.L020502} and YBa$_2$[Cu$_{1-y}$Zn$_y$]$_3$O$_{6+x}$ \cite{PhysRevLett.105.037207}. Such interesting intertwinement of the carrier doping and nonmagnetic substitution provides a rich playground to explore new physics in materials. 

Comparing to the mostly investigated cuprates, the perovskite iridate \SIO{} is naturally self-doped due to the Dirac semi-metal nature \cite{PhysRevLett.114.016401,liu2016direct,sen2020strange,PhysRevB.95.121102}. This makes the perovskite iridates a cleaner electron-doped Mott system to investigate the evolution of magnetic exchange interaction under spin dilution, without the complexity from introducing carriers and spin dilution through two different processes. With the itinerant carriers from self-doping and further spin dilution by substituting Ir with nonmagnetic Sn, \SIOx{} shows quite interesting physical behaviors. The parent compound \SIO{} is a paramagnetic (PM) semi-metal \cite{PhysRevB.86.085149,PhysRevB.95.121102}. The substitution with nonmagnetic Sn actually drives an order-by-disorder magnetic transition from paramagnetic to long-range AFM order \cite{PhysRevLett.117.176603,negishi2019contrasted,fujioka2018charge}, and the AFM order is robust beyond 50$\%$ Sn substitution \cite{PhysRevLett.117.176603}.

We have systematically investigated the magnetic excitation in \SIOx{} with resonant inelastic X-ray scattering (RIXS) from the PM semi-metal parent compound to heavily spin diluted while AFM ordered compound at $x$ = 0, 0.03, 0.06, 0.1 and 0.2. The experimental data is reported elsewhere \cite{li2026}. In this work, we focused on building a linear spin-wave theory (LSWT) model to extract the effective exchange interaction strength. For $x$ = 0, 0.03, 0.06 and 0.1, the extracted effective magnetic exchange strength follows a linear spin-dilution law, i.e. $J_{\mathrm{eff}}(x)=J_{0}(1-2x)$, where $J_0$ is the exchange interaction of the parent compound \SIO{}. Upon heavy substitution of $x$ = 0.2, $J_{\mathrm{eff}}$ deviates from linear scaling, likely due to clustering of the Sn dopants. These results not only reveal the robustness of the electron correlation effect in \SIOx{} and establish the parent compound \SIO{} as a strongly electron-correlated metal, but also provide microscopic information to understand the interplay of the charge and the spin dynamics in this system.   

\section{Experiment Results And Modeling}

\begin{figure}[h]
 \centering
 \includegraphics[width=.8\textwidth]{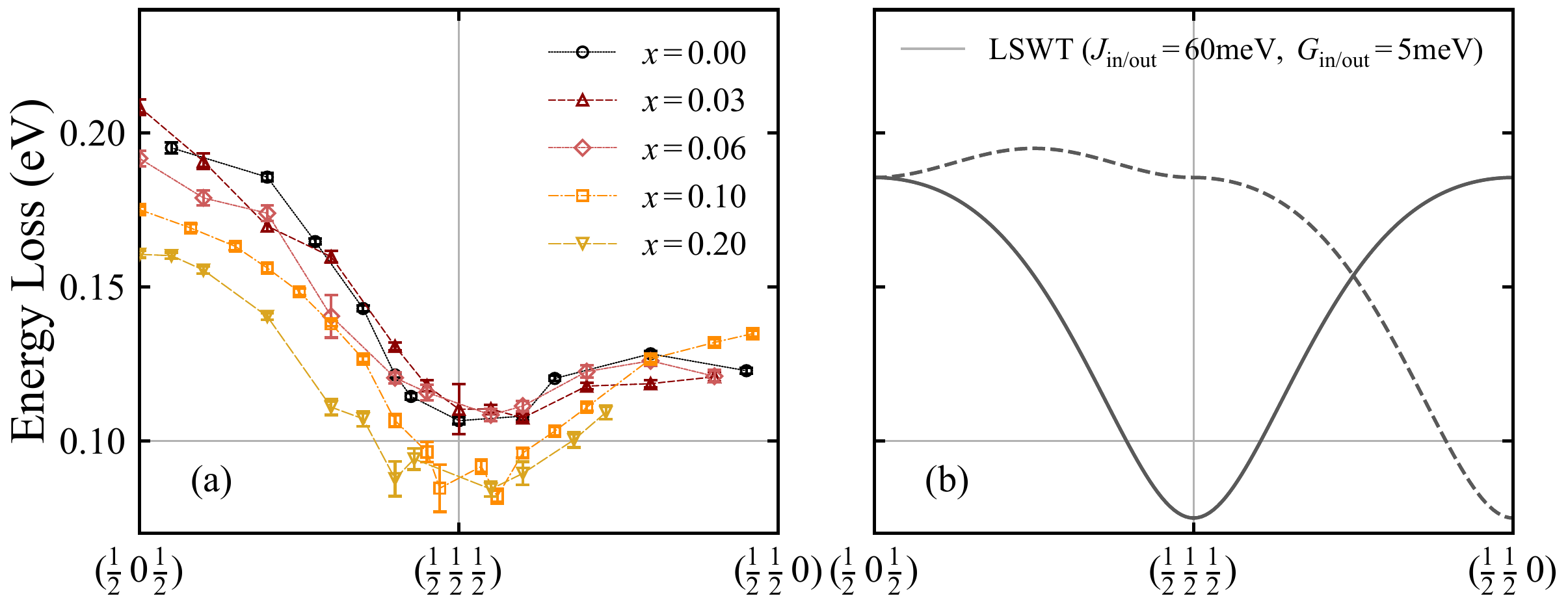}
 \caption{(a) Experimental dispersions of magnetic excitations of \SIOx{}. (b) Simulated dispersions by LSWT. The in-phase pair and out-of-phase pair are represented by the solid and dashed lines respectively.}
\label{Data&LSWT}
\end{figure}

\SIOx{} are of $Pbnm$ space group, a distorted perovskite structure with oxygen atom displacements. In our previous work \cite{bcp9-zg8f}, the anisotropic AFM structure of \SIOr{} has been determined to be of $G$-type with the magnetic moments along the $c$ direction. Since the magnetic atom Ir forms a cubic sublattice \cite{10.1063/1.2908879}, here we index the reciprocal space with $P4mmm$ symmetry of this sublattice for simplicity. Fig.\ref{Data&LSWT}(a) shows the dispersions of magnetic exchange interaction along the $K$ and $L$ directions, extracted from our RIXS data \cite{li2026}. The band gap is located at $Q_{\mathrm{gap}} =$ (1/2 1/2 1/2), corresponding to the AFM Brillouin zone center. The band top sits at $Q_{\mathrm{top}} = $ (1/2 0 1/2), which is the zone boundary along the $K$ direction. Upon Sn substitution, the magnetic excitation energies at both the band top and band gap systematically decrease, indicating a reduction of effective magnetic exchange interaction.

Based on the magnetic dispersions and the determined magnetic structure \cite{bcp9-zg8f}, we constructed a minimum exchange interaction model to extract the effective nearest-neighbor magnetic exchange interactions of \SIOx{}. As the $Pbnm$ space group of \SIOx{} has two independent symmetry operations: in-plane $b$-glide and out-of-plane $m$-reflection, the model contains the exchange interactions for both the in-plane bond along $b$ direction and out-of-plane bond along $c$ direction. Within the Hamiltonian, AFM Heisenberg interactions are included to reproduce the observed $G$-type AFM structure. Besides, to account for the observed anisotropy evidenced by the large gap, Ising terms are also included. The spin is set to be $S_z=1/2$ due to the singly occupied hole in $t_{2g}$ orbitals of the Ir-O octahedral \cite{PhysRevLett.101.076402,PhysRevLett.108.177003,PhysRevLett.109.157402}, and the spins are set to be along the crystallographic $c$ direction according to the determined magnetic structure. The Hamiltonian of this minimum model can be written as below:

\begin{equation*}
\begin{aligned}
 H_{\mathrm{in}}& =\sum_{\langle i,j\rangle}J^{ij}_{\mathrm{in}}(\vec{S}_i\cdot\vec{S}_j)+G^{ij}_{\mathrm{in}}S_i^zS_j^z
\\
 H_{\mathrm{out}}& =\sum_{\langle i,j\rangle}J^{ij}_{\mathrm{out}}(\vec{S}_i\cdot\vec{S}_j)+G^{ij}_{\mathrm{out}}S_i^zS_j^z
\end{aligned}
\end{equation*}

We applied LSWT to calculate the magnetic excitation dispersions \cite{PhysRevLett.126.087001}. The four nonequivalent Ir sites in the unit cell of \SIOx{} give rise to four magnon branches, which form two doubly degenerate pairs in our minimal model. These two pairs are distinguished by the in-phase and out-of-phase motions of spin with the same orientation. Since the RIXS cross-sections of the branch $n$ are given by $I_{n,Q}\propto \left |\sum_{i} \boldsymbol{M}_{n,Q}^{i}\right|^{2}$, where $\boldsymbol{M}_{n,Q}^{i}$ is the motion vector of the $i$th spin in the unit cell of \SIO{} \cite{PhysRevLett.126.087001}, these two pairs are of different scattering strength. For example, by setting $J_{\mathrm{in/out}}=60$ meV and $G_{\mathrm{in/out}}=5$ meV, the magnon dispersions and their RIXS spectral weights are calculated and the results are shown in Fig.\ref{Data&LSWT}(b). The in-phase pair has finite intensity in the RIXS measurements, whereas the scattering cross-section of the out-of-phase pair is zero. Therefore, we focus on the in-phase branches.

To quantify the simulated dispersion and compare it with the experimental data, we discuss the calculated energies at three high symmetry points, namely $Q_{\mathrm{gap}}$, $Q_{\mathrm{top}}$, and $Q_L=$ (1/2 1/2 0). At $Q_{\mathrm{gap}}$ and $Q_{\mathrm{top}}$, the energies are expressed as
\begin{equation*}
E_{\mathrm{gap}}^2
= (2J_{\mathrm{in}}+2G_{\mathrm{in}}+J_{\mathrm{out}}+G_{\mathrm{out}})^2 - (2J_{\mathrm{in}}+J_{\mathrm{out}})^2
\end{equation*}
and
\begin{equation*}
E_{\mathrm{top}}^2
= E_{\mathrm{gap}}^2 + 4J_{\mathrm{in}}J_{\mathrm{out}} + (2J_{\mathrm{in}})^2 
\end{equation*}
respectively. The energy at $Q_L$ is expressed as 
\begin{equation*}
E_{L}^2
= E_{\mathrm{gap}}^2 + 8J_{\mathrm{in}}J_{\mathrm{out}}
\end{equation*}
These expressions show that the anisotropic Ising terms $G$ open the band gap, and determine the band gap energy $E_{\mathrm{gap}}$ together with the mixed terms $G\cdot J$. Away from band gap, the isotropic exchange $J_{\mathrm{in}}$ dominants $E_{\mathrm{top}}$. The energy $E_{L}$ is determined analogously. In the experimental observations, the dispersion strength along $K$ direction is much stronger than along $L$ direction. Therefore, stronger $J_{\mathrm{in}}$ than $J_{\mathrm{out}}$ is expected. To further simplify this model, we confine $G_{\mathrm{out}}/G_{\mathrm{in}} = J_{\mathrm{out}}/J_{\mathrm{in}}$. Under this constraint, the three free parameters $J_{\mathrm{in}}$, $G_{\mathrm{in}}$, and $J_{\mathrm{out}}$ can be uniquely determined from the experimentally measured dispersions.

\section{Results}

\begin{figure}[h]
 \centering
 \includegraphics[width=.8\textwidth]{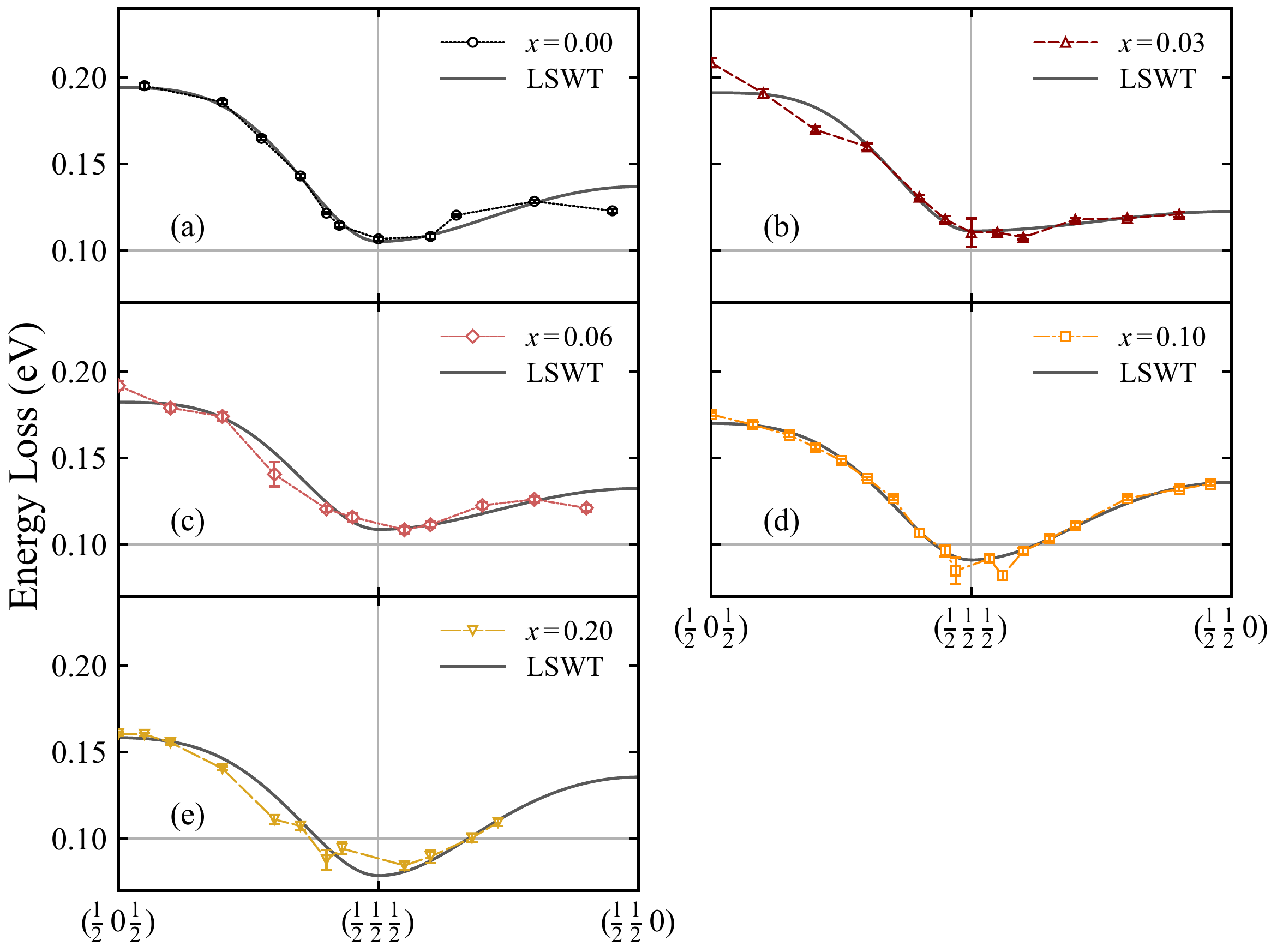}
 \caption{(a)-(e) LSWT fit to the experimental dispersions of \SIOx{} with $x$ = 0, 0.03, 0.06, 0.1 and 0.2.}
\label{fit}
\end{figure}

With above minimum model and LSWT, we fit the experimental dispersions of all spin-dilution ratios. The fitting results are shown in Fig.\ref{fit}, and the extracted exchange interaction parameters are listed in Table \ref{Jparameters}.

\begin{table}[h]
\caption{Effective magnetic exchange interactions (meV) of \SIOx{}.}
\centering
\begin{tabular}{ccccc}
		\hline
		\hline
		$x$&$J_{\mathrm{in}}$&$G_{\mathrm{in}}$&$J_{\mathrm{out}}$&$G_{\mathrm{out}}$\\ \hline
		0&75.5 ± 3.1&14.2 ± 1.4&12.7 ± 3.3& 2.4 ± 0.7 \\ 
		\hline 
        0.03&75.6 ± 2.8&17.3 ± 1.4&4.4 ± 1.7& 1.0 ± 0.4 \\ 
		\hline 
        0.06&68.1 ± 2.8&16.7 ± 1.2&10.4 ± 3.1& 2.6 ± 0.8 \\ 
		\hline 
        0.1&62.3 ± 1.1&11.2 ± 0.5&20.4 ± 1.0& 3.7 ± 0.3 \\ 
        \hline 
        0.2&56.5 ± 2.2&8.3 ± 2.1&27.0 ± 6.8& 3.9 ± 1.4 \\ 
		\hline 
		\hline
\end{tabular}
\label{Jparameters}
\end{table}

\begin{figure}[h]
 \centering
 \includegraphics[width=0.4\textwidth]{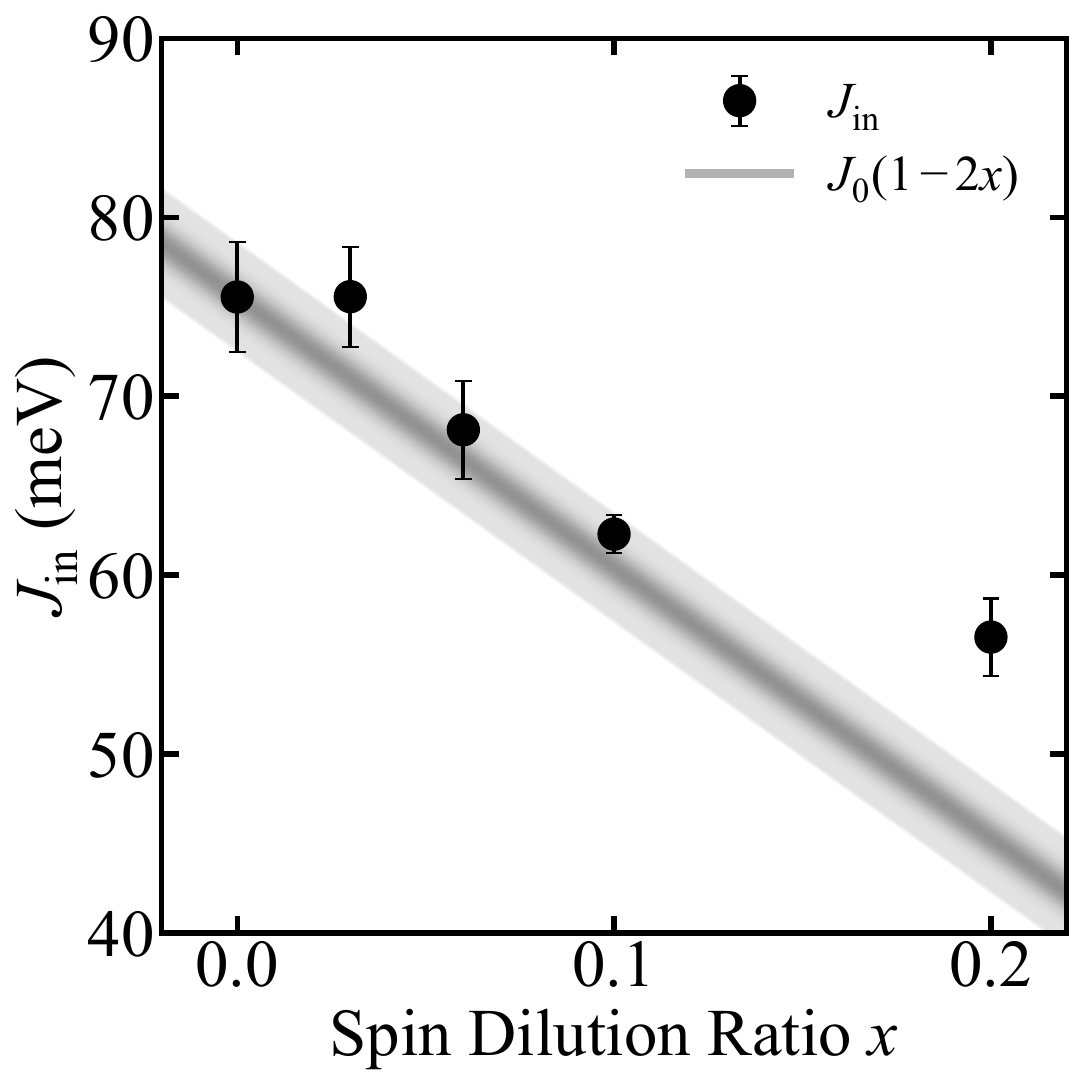}
 \caption{Evolution of the leading effective exchange interaction $J_{\mathrm{in}}$ with spin-dilution ratio $x$. The spin-dilution scaling law $J_{\mathrm{eff}}(x)=J_{0}(1-2x)$ is shown by the gray line.}
\label{doping}
\end{figure}

The doping evolution of dispersions are well described by the model. The gradually reduced dispersion energy upon Sn doping is reflected in the reduction of leading term $J_{\mathrm{in}}$ from 75.5 to 56.5 meV. We adopt a simple percolation model to understand the evolution of effective magnetic exchange interaction under spin dilution \cite{ulmke2007disorder,PhysRevB.58.8683}. For the in-plane Ir sublattice with $n\times n$ Ir sites, the number of in-plane exchange interaction bonds are $2n^2$ when $n$ is large. In the case with spin-dilution ratio $x$, $4n^2x$ bonds are broken at most. As a result, the ratio of the broken bonds are $2x$ at most. Thus the effective exchange interaction of \SIOx{} is expected to be $J_{\mathrm{eff}}(x)=J_{0}(1-2x)$ at least, with $J_0$ the in-plane exchange interaction of parent compound \SIO{} whose exchange paths are not disrupted. Fig.\ref{doping} shows the agreement of the evolution of the leading exchange term $J_{\mathrm{in}}$ to this simple broken-bond counting scenario. Upon spin-dilution, $J_{\mathrm{in}}$ indeed follows a linear behavior from $x$ = 0 to $x$ = 0.1. Further increasing the concentration to $x=0.2$, the effective $J_{\mathrm{in}}$ deviates from the linear scaling, likely due to the increased possibility of clustering of Sn dopants which induces less broken bonds.

\section{Discussion}

The \SIOx{} system is of interest with its peculiar phase diagram. The parent \SIO{} is a paramagnetic semi-metal, while spin-dilution with Sn drives a strong AFM insulator transition with the ordering temperature reaching $\sim 280$ K \cite{PhysRevLett.117.176603}. The metallic behavior of \SIO{} has been shown to be from the symmetry protected Dirac node in its topological band structure \cite{PhysRevLett.114.016401,sen2020strange}. Our results show that its average magnetic exchange interaction is linearly extrapolated from \SIOe{}, which is deeply in the AFM insulating phase with a transition temperature of $\sim 225$ K \cite{PhysRevLett.117.176603}. The agreement to the simple spin-dilution picture, where only broken exchange path counting is considered, clearly proves that the Dirac semi-metal \SIO{} hosts local electron-correlation strength and spin exchange interaction strength as strong as those in the \SIOx{} AFM insulators. The existence of highly mobile Dirac electrons likely introduces long range perturbation to the spin dynamics, thus suppresses an AFM transition. Locally, the spins are fluctuating with strong exchange interactions. 

In conclusion, we analyzed the dispersions of magnetic excitation in spin diluting \SIOx{} system measured by RIXS. The effective exchange interaction decreases with Sn concentration following the simple spin-dilution scaling law, $J_{\mathrm{eff}}(x)=J_{0}(1-2x)$. This behavior indicates the exist of strong electron-correlation effect in the Dirac semi-metal \SIO{}. We conclude the spin dilution disentangles the intertwined charge and spin dynamics and thus helps restore the long-range magnetic order, driving the order-by-disorder magnetic transition in \SIOx{}.

\funding{Xiang Li and Xuerong Liu were supported by the MOST of China under Grant No. 2022YFA1603900 and the startup fund from ShanghaiTech University. Y. J. was supported in part by the National Key R$\&$D Program of China under Grant No. 2022YFA1402703, NSFC under Grant No. 12347107 and 12574160. Y. W. was supported by the National Key R$\&$D Program of China under Grant No. 2022YFA1403800.}

\roles{Conceptualization and Investigation, X. Liu; Methodology, Y. J., Y. W., and X. Li; Writing, X. Liu and X. Li.}

\data{The data are available from the authors upon reasonable request.}

\bibliographystyle{iopart-num}
\bibliography{ref}
\end{document}